# Fe Intercalation under Graphene and Hexagonal Boron Nitride In-plane Heterostructure on Pt(111)

Igor Píš[a,b,*], Silvia Nappini[b], Federica Bondino[b], Tevfik Onur Menteş[a], Alessandro Sala[a,1], Andrea Locatelli[a], Elena Magnano[b,c,*]

[a]*Elettra - Sincrotrone Trieste S.C.p.A., S.S. 14-km 163.5, 34149 Basovizza, Trieste, Italy*
[b]*IOM-CNR, Laboratorio TASC, S.S. 14-km 163.5, 34149 Basovizza, Trieste, Italy*
[c]*Department of Physics, University of Johannesburg, PO Box 524, Auckland Park 2006, South Africa*

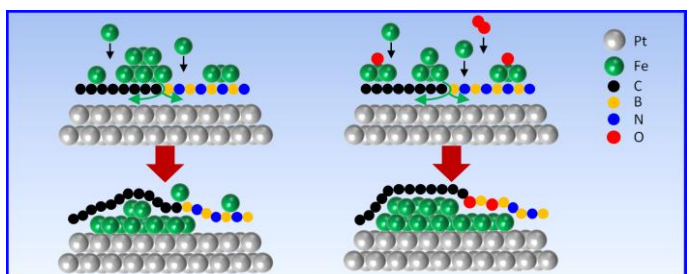

ABSTRACT: Metal nanostructures confined between sp$^2$ hybridized 2D materials and solid supports are attracting attention for their potential application in new nanotechnologies. Model studies under well-defined conditions are valuable for understanding the fundamental aspects of the phenomena under 2D covers. In this work we investigate the intercalation of iron atoms through a single layer of mixed graphene and hexagonal boron nitride on Pt(111) using a combination of spectroscopic and microscopic techniques. Thermally activated diffusion of iron proceeds preferentially under graphene and only partially under hexagonal boron nitride areas. When oxygen is coadsorbed with iron, the intercalation rate is higher, and formation of $B_2O_3$ and oxygenated B–C species is observed. Our results suggest the possibility of confining ferromagnetic layers under heterostructures of graphene and hexagonal boron nitride with potential technological implications in the fields of spintronics, magnetic data storage or chemistry under 2D covers.

---

[*] Corresponding authors. E-mail: igor.pis@elettra.eu (Igor Píš). E-mail: magnano@iom.cnr.it (Elena Magnano).
[1] Current address: University of Trieste and IOM-CNR, Laboratorio TASC, S.S. 14-km 163.5, 34149 Basovizza, Trieste, Italy

## 1. Introduction

Single layer graphene (Gr), hexagonal boron nitride (*h*-BN) and their monolayer heterostructures (*h*-BNG) supported on solid surfaces are expected to have outstanding electronic, mechanical, and chemical properties. In particular, the possibility to embed insulating and semimetallic materials within an atomically thin layer opens up the conception of new materials with great potential in nanotechnologies, such as novel electronic and spintronic devices [1–3] or 2D nanoreactors [4–7].

The properties of supported mixed h-BNG layers depend on their chemical and structural composition [8], but they can be modulated or significantly altered also by introduction of other species between the $sp^2$-bonded layers and their support [9]. For instance, hybridization of Gr $\pi$ states with intercalated-metal $d$ states can lift up the natural spin degeneracy of the Gr π-band [10–16]. Stacked heterostructures composed of ultrathin ferromagnetic (FM) films covered by Gr or graphene-like overlayers have captured the attention for their intriguing electronic and magnetic properties. Graphene can dramatically enhance magnetic anisotropy [17] or spin filtering effect [18] of the underlying FM thin film, while acting as a surface passivation oxidation-resistant layer at the same time [19]. 2D layers with $sp^2$ hybridization are also attracting attention in the emerging field of chemistry under cover. Interfaces between 2D $sp^2$ layers and substrate surfaces provide confined spaces for unusual chemical processes, different from their 3D counterparts. Atoms and molecules can be trapped at the interface, where they can further react [20–24].

Model studies under well-defined conditions are valuable for understanding the fundamental aspects of the new phenomena under 2D covers. Although several studies have been devoted to gas and metal intercalation under Gr [10,12,15,25–30] and *h*-BN [20,31–35], little research has been done on transition metal (TM) intercalation under mixed in-plane *h*-BNG heterostructures. Here, we examine the Fe intercalation into the *h*-BNG/Pt(111) system. The *h*-BNG interacts weakly with the Pt(111) substrate allowing intercalation and reaction of many molecular and atomic species. In addition, iron is a promising metal for growing tailored magnetic nanostructures on 2D hybrids of Gr and *h*-BN and their interfaces [36].

It is also desirable to identify and understand factors which considerably influence the rate of intercalation. It has been demonstrated that small amounts of coadsorbed oxygen can radically stimulate penetration of cobalt atoms through *h*-BN on Rh(111) [31]. Similar behaviour is expected also in case of Fe on the hybrid *h*-BNG. However, the presence of defects, such as, heteroatoms or boundaries between *h*-BN and Gr, can profoundly affect intercalation scenario [37].

The goal of the present work is to observe and understand basic processes and mechanisms accompanying thermally induced Fe intercalation under hybrid *h*-BNG single layer grown on Pt(111). In addition, the influence of a small amount of oxygen on the composite system is investigated and discussed.

Surface-sensitive experimental techniques are employed, namely, synchrotron radiation X-ray Photoelectron Spectroscopy (XPS), soft X-ray Absorption Spectroscopy (XAS), X-ray Photoemission Electron and Low Energy Electron Microscopy (XPEEM and LEEM, respectively). Their combination allows us to follow the structural and chemical composition changes.

2. **Experimental**

The measurements using synchrotron radiation were performed at the Elettra synchrotron facility in Trieste, Italy. The samples were characterized by means of high-resolution XPS and XAS at the CNR Beamline for Advanced diCHroism (BACH) [38,39]; LEEM and XPEEM measurements were carried out at the Nanospectroscopy beamline. The Nanospectroscopy beamline operates spectroscopic photoemission and low energy electron microscope (SPELEEM) and is equipped with an energy filter [40,41]. The instrument combines LEEM and XPEEM techniques, which provide structural and chemical sensitivity, respectively. In the SPELEEM, the electron kinetic energy is controlled by biasing the sample with a potential referred to a start voltage ($V_{start}$). The microscope lateral resolution approaches 10 nm in LEEM and 30 nm in XPEEM mode; energy resolution is about 0.3 eV in imaging spectroscopy.

Monolayer heterostructure of Gr and *h*-BN was prepared by pyrolytic decomposition of dimethylamine borane (DMAB; Sigma Aldrich, 97.0%) on a clean Pt(111) substrate kept at a temperature of 1000 K. DMAB was dosed in a separate UHV preparation chamber from a glass tube via a leak valve, keeping both at a constant temperature of 314 K. DMAB was purified by several freeze–pump–thaw cycles prior to the dose and the purity was checked by a quadrupole mass spectrometer. The Pt(111) crystal was exposed to 150 L of DMAB at a pressure of $6.65\times10^{-7}$ mbar to prepare complete *h*-BNG monolayer. This growth route leads to the formation of a continuous layer composed of Gr and *h*-BN domains with very small amount of B and N doped Gr and other BNC hybrid domains. Gr and *h*-BN comprised 71±7 and 27±6 at.% of the layer, respectively, as determined by quantitative XPS analysis. More details on the structure and composition of the *h*-BNG grown on Pt(111) can be found elsewhere [42,43].

The Pt(111) substrate (MaTecK, GmbH, 99.999% purity) was cleaned by cycles of 1.5 keV Ar$^+$ ion sputtering, annealing in $O_2$ (p($O_2$) = $2\times10^{-7}$ mbar) at about 900 K for a few minutes and annealing in UHV at >1000 K. The cleanliness and quality of the Pt(111) surface were checked by XPS and Low Energy Electron Diffraction (LEED). The sample temperature was measured by an N-type thermocouple clamped to the edge of the Pt single crystal at the BACH beamline, and a C-type thermocouple attached to the sample cartridge at the Nanospectroscopy beamline.

Iron metal was deposited from an iron rod heated by high-energy electron beam bombardment, while keeping the substrate at room temperature. An evaporation rate of 0.16 ML/min was calibrated in a separate experiment using the well-known reconstructions of $FeO_x$ ultrathin films on Pt(111) [44]. One monolayer (ML) of Fe is defined here as one Fe pseudomorphic layer, corresponding to $1.5\times10^{15}$ atoms/cm$^2$. The deposition was carried out in UHV conditions at a base pressure of $1\times10^{-9}$ mbar or in low-pressure oxygen ambience at $p(O_2) = 1.2\times10^{-8}$ mbar. Subsequent annealing at 600 and 700 K was performed in UHV, keeping the substrate at the selected temperature for 5–10 minutes. All additional exposures to oxygen were done at RT.

High-resolution XPS spectra were acquired by means of a VG-Scienta R3000 hemispherical electron energy analyser [45] at the BACH beamline. The peak intensities were normalised to the incident photon flux. Pt $4f_{7/2}$ and Fe 2p XPS spectra were measured at an angle of 60° with respect to the normal emission, the other XPS spectra were recorded in normal emission geometry. C 1s, N 1s, B 1s, and Pt $4f_{7/2}$ core levels were measured at a photon energy of 529 eV with a total energy resolution of 0.15 eV. Photon energies of 655 eV and 1030 eV with a resolution of 0.22 eV and 0.35 eV were employed for O 1s and Fe 2p spectra, respectively. The binding energy scale was calibrated with respect to the Fermi edge of the platinum substrate. Pt 4f and C 1s spectra were decomposed into spectral components using a Shirley-type background and Doniach-Šunjić line shapes convoluted with a Gaussian profile; other spectra were fitted with Voigt line shapes. Fe $L_{3,2}$ XAS spectra were acquired in total electron yield mode with a photon energy resolution set to 0.1 eV.

Sample *h*-BNG/Fe/Pt(111) prepared after Fe deposition and vacuum annealing at 800 K was transferred into the XPEEM–LEEM apparatus through air. Annealing to 625 K for 1 hour was applied prior to the measurements in order to remove molecular species adsorbed during the air exposure.

## 3. Results

### 3.1 *Thermally Induced Fe Intercalation into the h-BNG/Pt(111) Interface*

High-resolution XPS was used to follow the thermally activated Fe diffusion below the *h*-BNG single layer grown on Pt(111). C 1s, B 1s, N 1s and Pt $4f_{7/2}$ core level spectra were measured to characterize changes in the chemical and structural composition of the Fe–*h*-BNG–Pt(111) system. The measured spectra are reported in Fig. 1. Prior to the Fe deposition, the core levels (denoted as *h*-BNG) show spectral shapes and binding energies consistent with the experimental results reported earlier [42,43]. The carbon spectrum can be fitted with a narrow peak of Donjach-Šunjić shape centred at 283.9 eV, which corresponds to $sp^2$ hybridized carbon in Gr lattice, and a weak low-energy component assigned to B–C $sp^2$ and $sp^3$ bonds. Signs of B–C bonds are visible also in the B 1s spectrum at binding energies below 189 eV. The B 1s core level is dominated by an asymmetric peak decomposed into the main component centred at 189.7 eV and secondary component at 190.2 eV. This binding energy and shape are fingerprints of $sp^2$ hybridized boron in single layer *h*-BN on Pt(111) [34,46–48], similarly as the N 1s peaks at a binding energy of 397.4 eV and 398.0 eV. An additional weak N 1s component at 399.0 eV can be attributed to pyridinic nitrogen in Gr domains or at *h*-BN–Gr domain boundaries [49]. All spectra are consistent with a separate-phase scenario, where the great majority of *h*-BNG consists of large, adjacent patches of Gr and *h*-BN and Gr with low amount of chemical doping [42,43].

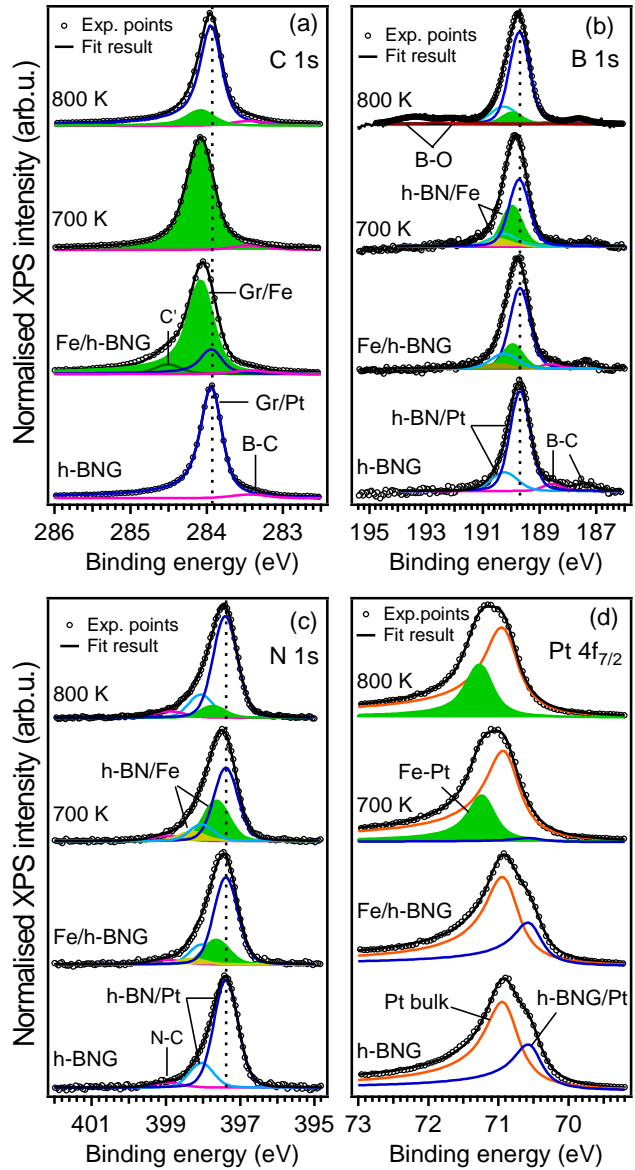

**Fig. 1.** XPS spectra (from bottom to top) measured after the *h*-BNG single layer growth on Pt(111); after 1 ML Fe deposition; and after subsequent vacuum annealing at indicated temperatures. The annealing at 800 K was preceded with 150 L $O_2$ dose at RT. The spectra were acquired with photon energy hν = 529 eV and normalised to their intensity maxima after background subtraction.

Pt $4f_{7/2}$ spectrum from the platinum substrate exhibits two main components after the *h*-BNG growth. The strongest one at 70.9 eV, characteristic for Pt metal, is assigned to bulk atoms. The second component originates from platinum atoms in the topmost layer. The shift of this component is sensitive to Pt coordination number and depends on the amount and nature of the atoms present above [50]. The measured core-level shift of approximately 0.4 eV towards lower binding energy is comparable with clean Pt(111) surface, confirming the weak interaction between the metal substrate and sp$^2$ hybridized *h*-BNG overlayer.

After the preparation of *h*-BNG/Pt(111) and its characterization, 1 ML of Fe was deposited at room temperature in UHV. The corresponding Fe $2p_{3/2}$ spectrum, reported in Fig. 2, has the shape and binding energy (706.8 eV) characteristic for metallic iron [22,51–53]. The ratio between Pt bulk and surface components of Pt $4f_{7/2}$ XPS spectrum (Fig. 1d) remains the same. The negligible changes in the Pt $4f_{7/2}$ surface component indicate that the iron remains on top of the *h*-BNG separated from the Pt topmost layer [22], most likely with morphology different from a flat layer [12]. The C 1s spectrum appears broader and shifted towards higher binding energy (Fig. 1a). The broadened peak can be well fitted with three components. The low binding energy component at 283.9 eV corresponds to pristine Gr, the dominant component at 284.1 eV is attributed to Gr weakly interacting with iron, and the small component C′ at 284.5 eV can be assigned to a stronger interacting Fe/Gr interface [30], n-doped Gr [54], or to $sp^3$ hybridized carbon induced by the presence of iron [55,56]. The detected presence of pristine Gr component after 1 ML Fe deposition is consistent with the expected Volmer-Weber 3D growth of Fe clusters on Gr [12].

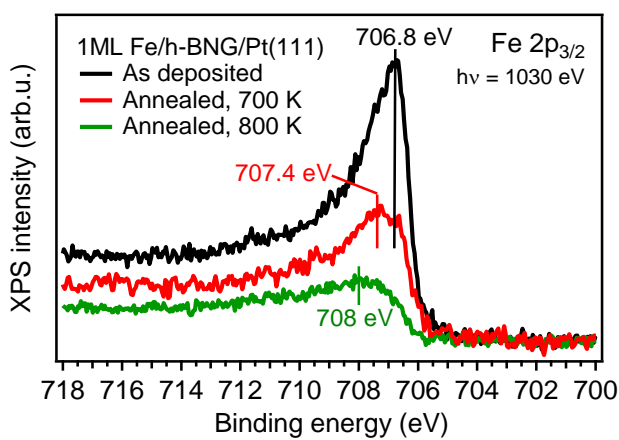

**Fig. 2.** Fe $2p_{3/2}$ XPS spectrum for 1 ML Fe deposited on h-BNG/Pt(111) at RT and after annealing at 700 K. The sample was then exposed to 150 L $O_2$ at RT and annealed at 800 K subsequently.

The Fe deposition at RT has only a small impact on nitrogen and boron spectra. A small shift towards higher binding energy is observed in both N 1s and B 1s core levels. The best fits of B 1s and N 1s spectra are obtained assuming new components at 397.6 eV in N 1s and 190.0 eV in B 1s spectrum and their high binding energy satellites with the same relative shifts and intensities as for the pristine *h*-BN (Fig. 1b, c). The new components can be attributed to the B and N atoms in *h*-BN lattice underneath Fe metal islands. Noteworthy, after the iron deposition, preferential attenuation of C 1s signal compared to B 1s and N 1s signals is observed (Fig. 3a). This difference indicates higher concentration of Fe on top of the Gr domains. However, morphology effects related to different Fe island size on Gr and *h*-BN could also contribute to the observed preferential C 1s signal attenuation.

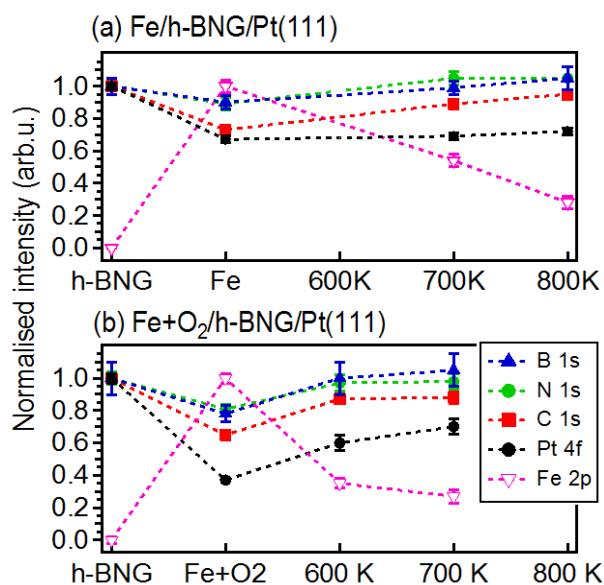

**Fig. 3.** XPS integrated intensities measured after the h-BNG growth; (a) 1 ML Fe deposition; annealing at 700 K; and 800 K; (b) 1 ML Fe deposition in low-pressure $O_2$; annealing at 600 K; and 700 K. The intensities are normalised to their initial values after the h-BNG growth (B 1s, N 1s, C 1s, Pt 4f) and Fe deposition (Fe 2p).

After the deposition, the sample was annealed at 700 K for 5 minutes. As reported previously [12,23,26], such conditions induce complete Fe intercalation at the interface between Gr and Pt(111), but prevent formation of Fe-Pt alloy (see also Fig. S1 in the supplementary information (SI)). Signs of Fe diffusion under the *h*-BNG could be observed also here. First, C 1s, N 1s and B 1s XPS spectra are shifted to higher binding energies after the annealing (Fig. 1; dashed lines). Second, the carbon, nitrogen and boron signals are more intense (Fig. 3a), while the Fe $2p_{3/2}$ XPS intensity is reduced (Fig. 2), as the signal from Fe atoms is now attenuated by the above lying *h*-BNG. Iron desorption can be neglected considering the very low Fe vapour pressure at the used temperature. At the same time a new component around 707.4 eV appears in the Fe $2p_{3/2}$ XPS spectrum. This positive binding energy shift can be ascribed to the transformation from Fe–Fe to Fe–Pt coordination when Fe atoms are located on top of Pt(111) [57]. Presence of iron between the *h*-BNG and Pt substrate is further confirmed by the reduction of the Pt $4f_{7/2}$ low energy component from the weakly interacting Pt superficial atoms and appearance of new component at 71.2 eV (Fig. 1d) ascribed to Fe–Pt bond [26,53]. The C 1s spectrum shows only Gr/Fe component, meaning that Fe is intercalated under all Gr domains. Compared to other Gr/Fe/TM systems [12,29,30], the relatively low C 1s binding energy of 284.1 eV can be assigned to Pt–Fe interaction and a minor intermixing [30,58]. More intense are also the *h*-BN/Fe components in both N 1s and B 1s spectra (Fig. 1b, c). However, a significant part of the B and N signal is still coming from the pristine *h*-BN on Pt(111). Signal of *h*-BN/Pt(111) interface is present also in the Pt 4f spectrum, where the low binding energy component is still detectable. The low intensity of this component is in accord with expectations, considering that *h*-BN is the minor component of *h*-BNG. On the bases of all the PES spectra presented in Fig. 1–3, we can conclude that Fe preferentially nucleates on and intercalates under Gr domains and only partially under *h*-BN, leaving some fractions of pristine *h*-BN/Pt(111).

It has been previously reported that Fe intercalated below Gr is well prevented from oxidation [12,23]. Oxygen resistance of the prepared *h*-BNG/Fe/Pt system after annealing at 700 K was probed by exposure to molecular oxygen (p(O$_2$) = 6.65×10$^{-7}$ mbar, 150 L) at room temperature. Corresponding Fe L$_{3,2}$ X-ray absorption spectra before and after the O$_2$ dose are plotted in Fig. 4a, b. A new spectral feature, corresponding to Fe$^{x+}$ oxides [12,59], appears at 709 eV. In contrast to the Fe/Gr/Pt(111) case, Fe/h-BNG/Pt(111) annealed at 700 K is still reactive towards low oxygen exposures. Subsequent annealing at 800 K for 5 minutes reduces the oxidized iron back to metallic state (Fig. 4c) and some oxygenated species of boron appear, as shown in Fig. 1d. The nature of B–O species will be discussed in the next section. No evidence of oxidation is observed after the further exposure to oxygen (Fig. 4d), meaning that all the Fe atoms are now protected against oxidation. On the other hand, the temperature of 800 K is leading to iron diffusion into the Pt substrate, as can be deduced from XPS spectra in Fig. 1–3. The C 1s, N 1s and B 1s spectra are shifted back towards binding energy values before Fe deposition pointing to Fe migration into Pt substrate layers and formation of h-BNG/Fe–Pt interface. It has been shown that the Pt concentration of about 50% in a bimetallic surface alloy can result in complete detachment of Gr from the substrate [58]. The conclusion about Fe–Pt alloying is further supported by Fe 2p$_{3/2}$ peak shift to 708 eV (Fig. 2), which is characteristic for Fe atoms embedded in the topmost Pt(111) layer [53,57].

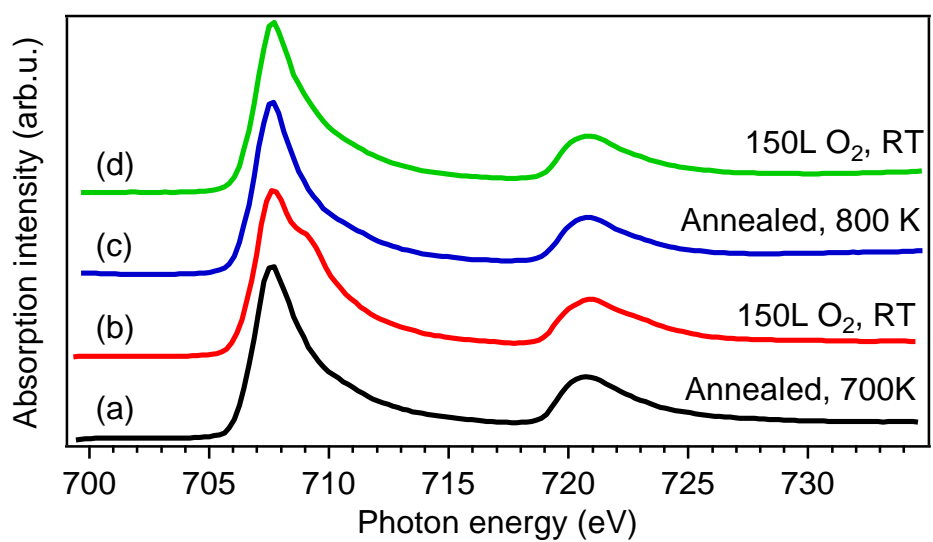

**Fig. 4.** Fe L$_{3,2}$ XAS spectrum measured after 1 ML Fe deposition on top of *h*-BNG/Pt(111) at RT followed by annealing at 700 K (a); exposure to 150 L O$_2$ (p(O$_2$)=6.65×10$^{-7}$ mbar) at RT (b); annealing at 800 K (c); and repeated exposure to O$_2$ (d). The spectra are normalised to maximum.

LEEM and XPEEM microscopic images of the sample after annealing at 800 K are presented in Fig. 5. The LEEM image (Fig. 5a) shows a clear contrast between Gr and h-BN areas due to a different electron reflectivity at low energy, as reported earlier [42]. Further evidence of the presence of laterally separated h-BN and Gr regions, which cover the surface in a continuous manner, is given by the complementary C 1s (Fig. 5b), N 1s, B 1s XPEEM images and microspot electron energy loss spectra (µ-EELS), all available in Fig. S2 in the SI.

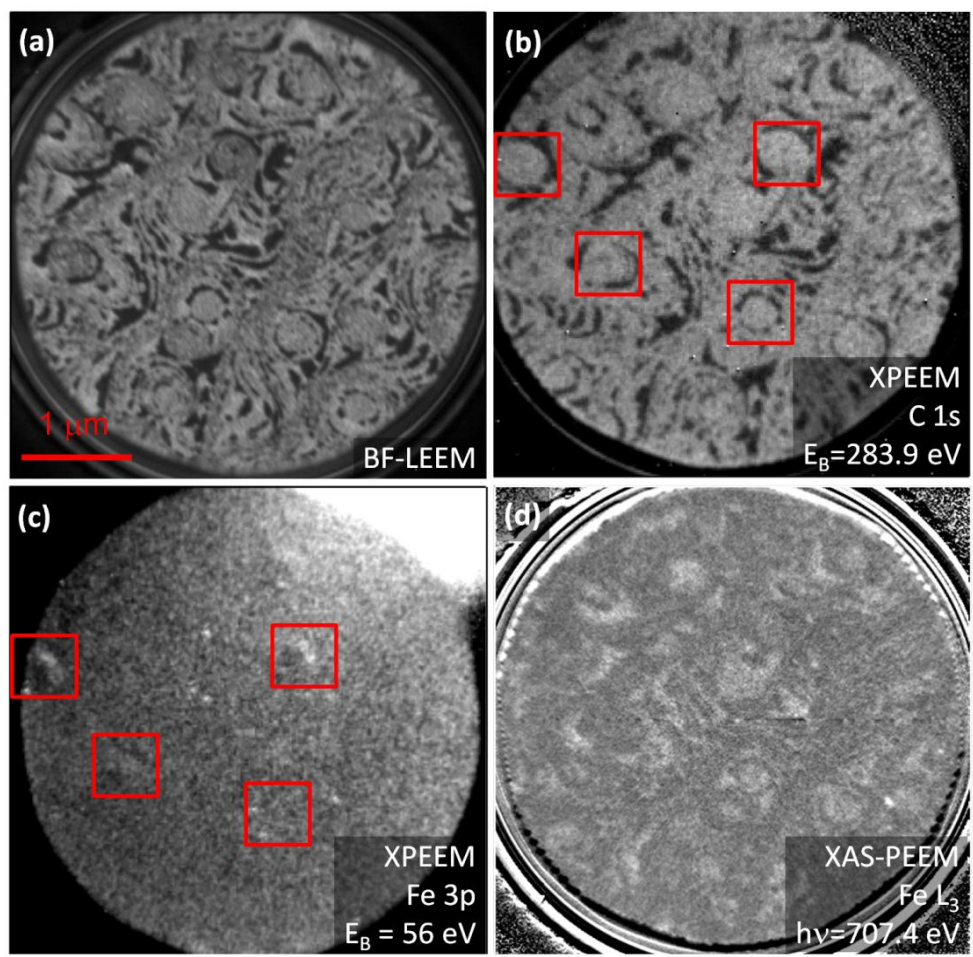

**Fig. 5.** LEEM and element specific microscopic XPEEM and XAS-PEEM images obtained after vacuum vapour deposition of 1 ML Fe on top of h-BNG/Pt(111) and subsequent annealing at 800 K for 5 minutes. (a) Brighter and darker areas in bright field (BF) LEEM ($V_{start}$ = 6 V) correspond to Gr and h-BN, respectively. Photon energies of 400 eV and 160 eV were employed for (b) C 1s and (c) Fe 3p, respectively. (d) XAS-PEEM image measured with secondary electrons ($V_{start}$ = 1.5 V) under 707.4 eV X-ray beam, corresponding to Fe $L_3$ absorption edge. The areas marked by squares indicate positions with higher Fe 3p signal variation.

The lateral distribution of Fe was recorded by core-level XPEEM at Fe 3p line (Fig. 5c) and X-ray Absorption Spectroscopy PEEM (XAS-PEEM) at Fe $L_3$ edge. The Fe 3p XPEEM intensity map, taken with an electron kinetic energy of 112 eV to achieve high surface sensitivity, exhibits a low contrast implying similar amount of Fe under Gr and h-BN after 800 K annealing. Nevertheless, it is possible to recognize a few areas that exhibit different behaviour. Their positions correspond to Gr islands on top of flat circular terraces surrounded by step bunches (Fig. 5b, c). The higher Fe signal variation is then ascribable more to morphological effects, in particular to substrate step bunches, which are preferential channels for Fe intercalation and bulk diffusion [60]. On the other hand, the larger probing depth provided by XAS-PEEM proves that a higher concentration of Fe is detectable under Gr than under $h$-BN. The bright areas with higher near-surface Fe concentration match well carbon areas, which exhibit higher intensities in the C 1s XPEEM image. The XPEEM and XAS-PEEM images thus confirm the conclusions driven from XPS: Fe is preferentially embedded between Pt and Gr domains and interdiffuses in the Pt substrate after annealing at 800 K. In addition, the microscopic images also show that the intercalation is uneven, as the highest Fe concentration is observed only under some of the Gr islands.

### 3.2 *Influence of Coadsorbed Oxygen*

In order to examine the influence of oxygen surfactant on Fe intercalation, the previous experiment was repeated with Fe vapour deposition in low-pressure oxygen atmosphere ($p(O_2)$ = $1\times10^{-8}$ mbar). Overall, the sample was exposed to approximately 3 L of $O_2$ during the deposition. The XPS spectrum of iron just after the deposition (Fig. 6) shows the Fe $2p_{3/2}$ maximum at 706.8 eV and a pronounced shoulder around 710 eV. While the former component corresponds to metallic iron, the latter one is characteristic for $Fe^{2+}$ charge state and it is attributed to FeO oxide [22,52,61]. However, a certain amount of $Fe^{3+}$ with corresponding component at 711 eV cannot be excluded because of the broad shape of the high-energy shoulder. O 1s spectrum reported in Fig. 7 shows the dominant peak at 529.9 eV attributed to iron oxides [51]. Second component at 531.5 eV could be ascribed to subnanometer FeO clusters [62] or Fe–OH [63]. The oxygen spectrum also contains a weak high energy component at 532.9 eV assigned to B–O species, as will be discussed later. It should be noted here that no oxidation has been detected after similar $h$-BNG/Pt(111) exposure to $O_2$ (see Fig. S3 in the SI). Relative concentrations of iron and oxygen estimated from the XPS intensities indicate roughly one adsorbed oxygen atom per each three iron atoms. However, the iron to oxygen ratio can differ locally and pure metal or mixed metal/oxide islands could be formed.

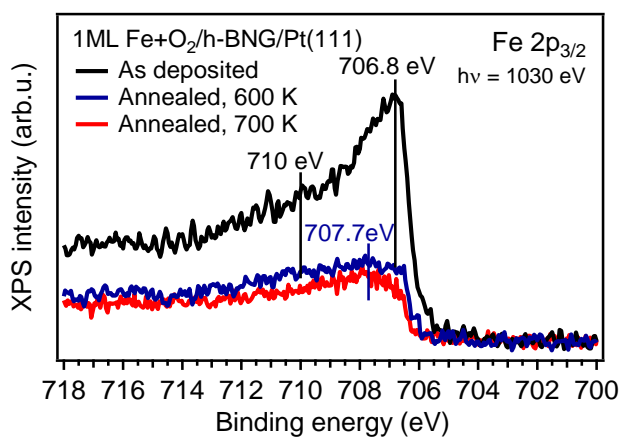

**Fig. 6.** Fe $2p_{3/2}$ XPS spectra for 1 ML Fe deposited on h-BNG/Pt(111) at RT in low-pressure oxygen atmosphere (p($O_2$) = $1.2\times10^{-8}$ mbar) and after subsequent vacuum annealing at the indicated temperatures.

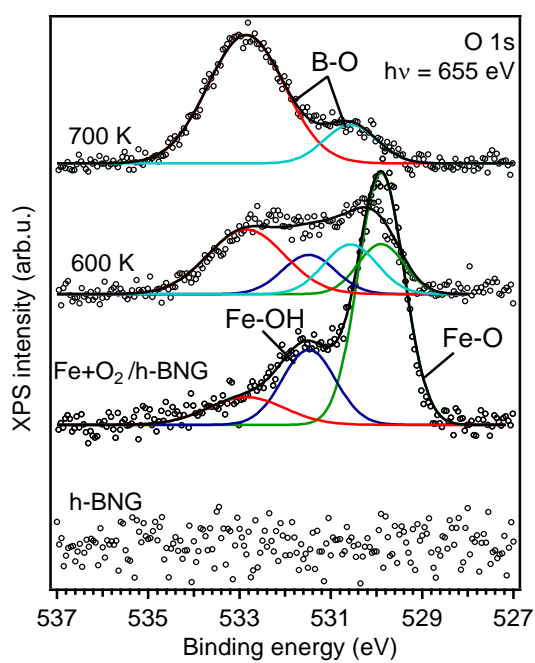

**Fig. 7.** O 1s XPS spectra measured after 1 ML Fe deposition at RT in low-pressure $O_2$ atmosphere (p($O_2$) = $1.2\times10^{-8}$ mbar) and after subsequent vacuum annealing at the indicated temperatures. The spectrum acquired for h-BNG/Pt(111) prior to the deposition is also shown for comparison.

XPS signals of C, B, N and Pt (Fig. 8) exhibit only small differences in the spectral shapes with respect to those prior to the deposition, although a drop in intensity is evident (Fig. 3b). Preserved Pt $4f_{7/2}$ spectral shape (Fig. 8d) indicates that iron and oxygen are on top of the *h*-BNG layer.

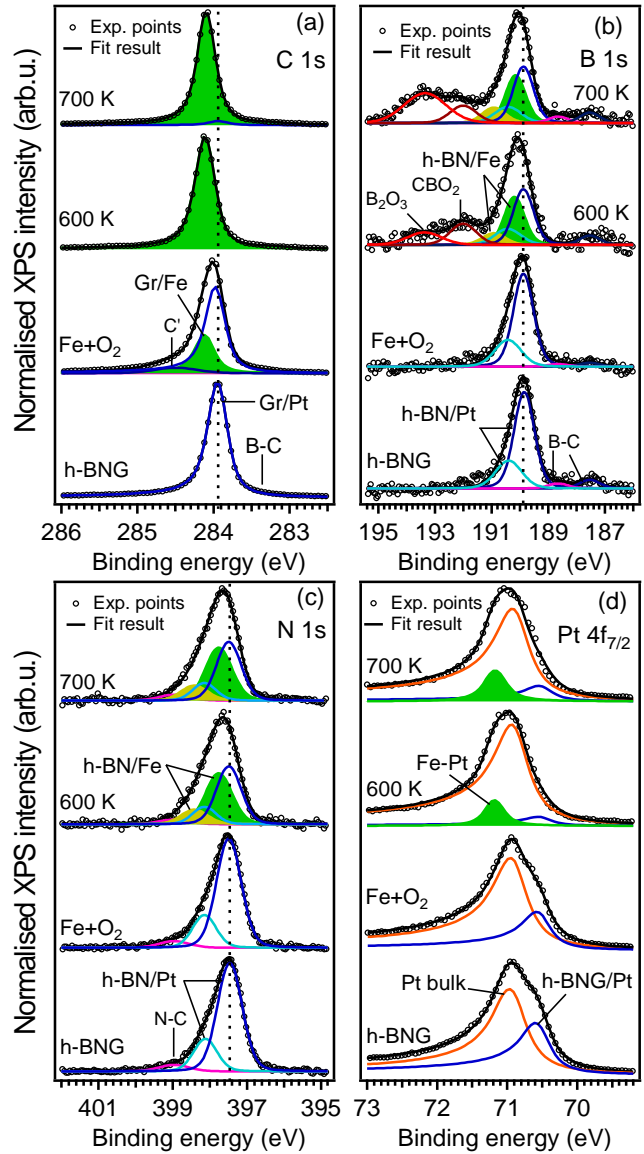

**Fig. 8.** XPS spectra (from bottom to top) measured after the h-BNG single layer growth on Pt(111); after the deposition of 1 ML Fe in $O_2$ ($p(O_2) = 1.2 \times 10^{-8}$ mbar); and after subsequent vacuum annealing at indicated temperatures. The spectra were acquired with photon energy $h\nu$ = 529 eV and normalised to their intensity maxima after background subtraction.

The absence of Fe/*h*-BN components in B 1s and N 1s spectra after the deposition at RT (Fig. 8b, c) could be ascribed to $FeO_x$ phase and weak bonds between $FeO_x$ and *h*-BN. The weak $FeO_x$/substrate chemical interaction can also explain the dominance of pristine Gr component in C 1s spectrum in Fig. 8a. Nonetheless, the C 1s components corresponding to Gr/Fe are still detectable, meaning that non-oxidized iron is also present on top of Gr domains.

After annealing at 600 K, the Gr/Fe component is dominant in the C 1s spectrum. Simultaneous N 1s and B 1s shift to higher binding energies, Fe $2p_{3/2}$ XPS intensity decrease (Fig. 6) and the appearance of Fe–Pt components in both Fe $2p_{3/2}$ (peak at 707.7 eV in Fig. 6) and Pt $4f_{7/2}$ spectra (peak at 70.2 eV in Fig. 8d) means that most of the iron is located between *h*-BNG and Pt(111). Compared with Fe signal measured after annealing to 700 K in the previous experiment (Fig. 3a), the Fe signal decrease in the case of coadsorbed oxygen is more pronounced already at 600 K (Fig. 3b). As this effect can be hardly ascribed to Fe diffusion into the Pt substrate or to Fe desorption at this temperature, it could be interpreted as more complete intercalation with respect to intercalation without oxygen. On the other hand, the relatively high Fe signal reduction and simultaneous increase of Pt substrate intensity (Fig. 3b) indicate that morphological changes, such as, formation of more than one ML thick Fe or $FeO_x$ clusters occur, too. While traces of FeO are still visible in Fe $2p_{3/2}$ spectrum after 600 K (Fig. 6), metallic component at 707.7 eV, corresponding to Fe located between $h$-BNG and Pt, dominates the spectrum after 700 K annealing.

The appearance of $h$-BN/Fe components in N 1s and B 1s spectra after annealing reveals the presence of metallic iron not only under Gr but also under the $h$-BN domains. Moreover, the intensity of the $h$-BN/Fe components indicates that the amount of intercalated Fe is higher compared to the iron deposition without any oxygen. More new features appear in B 1s. All of them can be attributed to oxygen containing B–O species. The highest binding energy component at 193.4 eV is assigned to the fully oxidized boron, $B_2O_3$ [64–66]. The component at 192.0 eV can be attributed to boron bonded to carbon and two oxygen atoms ($CBO_2$) [64,67–69]. The evidence of boron oxidation is present also in O 1s spectra (Fig. 7). While Fe–O features are vanishing upon annealing, new components are appearing at higher binding energy. After annealing at 700 K, $B_2O_3$ peak at 532.9 eV [31,66,68] is dominant. The other new O 1s component may be ascribed to the C–B–O species. The gradual transition from C–B–O to $B_2O_3$ indicates that oxidation starts at the Gr/$h$-BN boundaries. The total area of oxygen O 1s spectrum is lowered to about 67% of the initial value after first annealing and then it remains stable, meaning that 1/3 of the adsorbed oxygen is desorbed during the first annealing step and the rest is transformed from iron oxide to boron oxide species.

Like in the first section, oxidation resistance of $h$-BNG/Fe/Pt system after annealing at 700 K was probed by exposure to oxygen and subsequent Fe $L_{3,2}$ XAS spectroscopy (see Fig. S4 in the SI). In contrast to the previous case, no signs of Fe oxidation are observed, indicating completed Fe intercalation at 700 K.

4. Discussion

The intercalation of 1 ML Fe between $h$-BNG and Pt(111) was performed by annealing of RT-deposited Fe islands. It has been previously demonstrated that the thermal treatment at a temperature of 600–700 K results in complete Fe diffusion in between single layer Gr and Pt(111) substrate and morphology transition from small nanoparticles to atomically flat Fe layer. Furthermore, the intercalated iron was found to be remarkably protected from the interaction with the environment under ambient conditions [12,23]. Our results show that the situation is different in the case of *h*-BNG layer. Although significant amount of iron is found under the *h*-BNG after the annealing, the iron is not protected from oxidation, not even at mild ambient conditions. The observed oxidation could be due to the presence of some iron on the surface, as a result of a lowered intercalation rate. Another reason explaining the Fe oxidation could be $O_2$ intercalation. Although no signs of $O_2$ diffusion between *h*-BNG and Pt(111) is observed under the studied conditions, the presence of Fe, for example at or under Gr/*h*-BN boundaries, could open up new intercalation channels for oxygen.

The XPS spectra show that after annealing at 700 K Fe is present under all Gr, but only under some parts of *h*-BN. Similar different intercalation rate was reported also for Mn atoms on *h*-BN/Rh(111) and Gr/Rh(111) [33]. The inhibited intercalation under the *h*-BN domains can be related either to a higher Fe diffusion barrier or to a thermodynamically unfavourable configuration of Fe at the *h*-BN/Pt interface rather than in between Gr and Pt(111).

The obtained experimental results allow us to get also insight into the intercalation mechanism. On the basis of numerous studies of metal atom intercalation under a monolayer Gr, two main intercalation mechanisms have been proposed [4]. In the so-called *defect-aided intercalation* process, intercalation channels are the extended defects such as domain boundaries and point defects. In the *exchange intercalation* mechanism, on the other hand, the metal atoms are supposed to create transition-state defects in the Gr lattice, through which the intercalant atoms penetrate and the C–C bonds of the Gr lattice are reestablished afterwards [28]. The former mechanism has been suggested for elements which interact with carbon weakly or are large in their size, whereas the latter has been considered for strongly interacting elements, such as Fe.

Indeed, signs of metal-generated defects in Gr and their self-healing can be observed in the C 1s spectra (Fig. 1a, Fig. 8a), where the high binding energy component C' at 284.5 eV repeatedly appears after Fe deposition and disappears after thermally induced Fe intercalation. However, it was previously showed that iron nanoislands, vapour deposited on Gr/Pt(111) at RT, nucleate at contiguous Gr flake boundaries and point defects [12]. In the current study we observe transformation of Fe–O to C–B–O species after exposure to oxygen and annealing, indicating iron is situated nearby Gr/*h*-BN boundaries. In addition, XPEEM images showed that the metal substrate structure, such as step bunches have also influence on the iron migration. Thus, we suggest that also the *defect-aided intercalation* mechanism significantly contributes to the migration of Fe in between *h*-BNG and Pt(111).

The experiment with Fe deposition in oxygen atmosphere showed that iron intercalation can be significantly influenced by the presence of oxygen. During the Fe reactive deposition, some oxygen is dissociatively adsorbed and reacts with the iron. The Fe $2p_{3/2}$ XPS spectra show that among the possible oxide species $Fe^{2+}$ oxidation state is prevailing, indicating the formation of FeO. The formation of $Fe_3O_4$ could be expected as it is thermodynamically the most stable phase [52] and $FeO_{x+1}$ (x > 0) was indeed observed, for example, on Gr/Ni(111) and HOPG substrates [29,70]. However, we observe a low relative oxygen concentration after the used $O_2$ dose (Fe:O = 3:1 at.%) and the formation of $Fe^{2+}$ corresponding to FeO is typical of the initial stage of iron oxidation [71].

In the presence of oxygen, annealing at 700 K is sufficient to complete the iron intercalation. The small amount of coadsorbed oxygen evidently promotes the penetration of Fe through the *h*-BNG layer. Similar effect has been observed for cobalt atoms on *h*-BN/Rh(111) [31]. The adsorbed oxygen acted as a wetting agent suppressing formation of bigger Co clusters, weakening the metal–metal bond in individual nanoparticles, thus facilitating the intercalation at step edges and point defects. Similar mechanism could apply also for Fe on *h*-BNG. Unlike Gr, some *h*-BN parts without Fe underneath are observed after annealing at 700 K. Although the oxygen promotes the Fe diffusion under both *h*-BN and Gr domains, the migration under Gr is still more facile. The preferential migration under Gr and the presence of bare *h*-BN/Pt(111) areas implies that some of the Fe islands under the Gr may be more than 1 ML thick. Iron nanoparticle sintering under graphene is indicated also by Fe 2p intensity decrease (Fig. 2,3,6), in contrast to intensity increase expected for structural transformation from supported 3D nanoparticles to 2D intercalated interlayer [22,72].

In here studied case of Fe–O/*h*-BNG, about 34% (25%) of the boron is bonded to oxygen after the annealing to 700 K (600 K). This transition is in accord with the case of oxygen assisted Co intercalation [31], where oxygen migration from Co to *h*-BN was reported, although the boron oxidation was less pronounced. At variance with pure *h*-BN or Gr layers, new classes of defects, such as *h*-BN–Gr boundaries and N and B heteroatoms in Gr lattice are present in the *h*-BNG heterostructure. These defects together with the Fe at the boundaries between *h*-BN and Gr domains might be responsible for the substantial oxidation of the *h*-BN phase. Here observed oxidation of boron is apparently facilitated by the iron, on which molecular oxygen chemisorbs dissociatively. Thermally-induced reduction of iron oxides generates active atomic oxygen, which can participate in reactions with boron.

Oxygenated boron species, such as $CBO_2$, are observed after annealing to 600 K and $B_2O_3$ at higher temperature. The formation of carbon–boron–oxygen compounds implies that the boron oxidation and therefore also iron reduction might start at the *h*-BN–Gr boundaries. Noteworthy are the oxygenated boron species, which were showed to be catalytically active in oxygen reduction reaction with enhanced selectivity towards formation of $H_2O_2$ [7,67,73]. It is also worth to mention that no formation of oxygenated boron species has been observed on *h*-BN monolayers grown on TM surfaces, even at much higher oxygen pressures and comparable temperature [34,74].

## 5. Conclusions

We have investigated the process of thermally-stimulated Fe intercalation under a single layer *h*-BNG heterostructure prepared on Pt(111) and the influence of coadsorbed oxygen. Compared with pure single Gr layer, *h*-BNG layer requires higher temperature to complete the iron diffusion. Upon annealing, iron migrates and agglomerates preferentially between Pt(111) and Gr domains and only partially between *h*-BN and Pt(111).

Oxygen coadsorbed with iron during the deposition promotes the intercalation. A temperature of 700 K is found sufficient to intercalate 1 ML Fe deposited in low-pressure oxygen atmosphere ($p(O_2) = 1 \times 10^{-8}$ mbar). The oxygen bound to iron is released upon annealing, reacts with boron in *h*-BNG and forms $B_2O_3$ and C–B–O functional groups.

Our results indicate that thin Fe layers and compounds of particular configuration can be grown and stabilized under *h*-BNG 2D cover. This finding opens up the opportunity to study their electronic, magnetic, as well as catalytic properties. Fe-based nanocatalysts covered with $sp^2$ hybridized shell may exhibit intriguing activity or selectivity in many chemical reactions.

**Acknowledgements**

This work was supported by Consiglio Nazionale delle Ricerche (CNR) and the Italian Ministero dell'Istruzione, dell'Università e della Ricerca (MIUR) through the national grant Futuro in Ricerca 2012 RBFR128BEC "Beyond graphene: tailored C-layers for novel catalytic materials and green chemistry" and by MIUR through the program "Progetto Premiale 2012"–Project ABNANOTECH. Federico Salvador, Davide Benedetti, Aleksander De Luisa and Paolo Bertoch are acknowledged for technical support.

**Appendix A. Supplementary data**

Additional XPEEM images, µ-EELS, XPS and XAS spectra.

# Supplementary Information

# Fe Intercalation under Graphene and Hexagonal Boron Nitride In-plane Heterostructure on Pt(111)

Igor Píš [a,b], Silvia Nappini[b], Federica Bondino[b], Tevfik Onur Menteş[a], Alessandro Sala[a], Andrea Locatelli[a], Elena Magnano[b,c]

[a]Elettra - Sincrotrone Trieste S.C.p.A., 34149 Basovizza, Trieste, Italy
[b]IOM-CNR, Laboratorio TASC, 34149 Basovizza, Trieste, Italy [c]Department of Physics, University of Johannesburg, Auckland Park 2006, South Africa

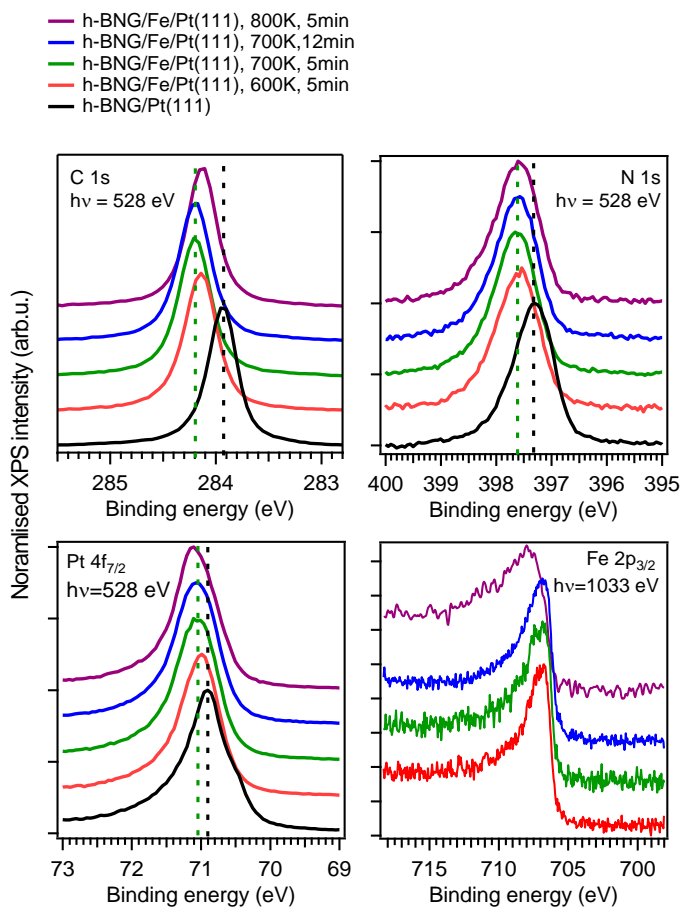

**Fig. S1.** Photoemission core levels of *h*-BNG/Fe/Pt(111) and after 1 ML Fe deposition at RT and subsequent annealing at the indicated temperatures for the indicated periods of time. Spectra for *h*-BNG/Pt(111) are also shonwn for comparison. The C 1s signals from graphene and N 1s from *h*-BN reach maximum shift (indicated by the dashed lines) to higher binding energy at 700 K, which corresponds to Fe intercalation between *h*-BNG and Pt(111). The intercalated Fe layer is stable at 700 K for at least 12 minutes. Clear spectroscopic signs of Fe–Pt alloying are observable after annealing at 800 K for 5 minutes. C 1s and N 1s peaks are shifted to lower energies, while Pt $4f_{7/2}$ and Fe $2p_{3/2}$ to higher energies. All spectra are normalised to intensity maxima. Fe $2p_{3/2}$ spectra were recorded at an emission angle of 60°, the others at normal emission geometry. Note that the Pt $4f_{7/2}$ spectra reported in the main text of the article were acquired at the emission angle of 60° at which the relative intensity of the Pt $4f_{7/2}$ surface component is profoundly enhanced.

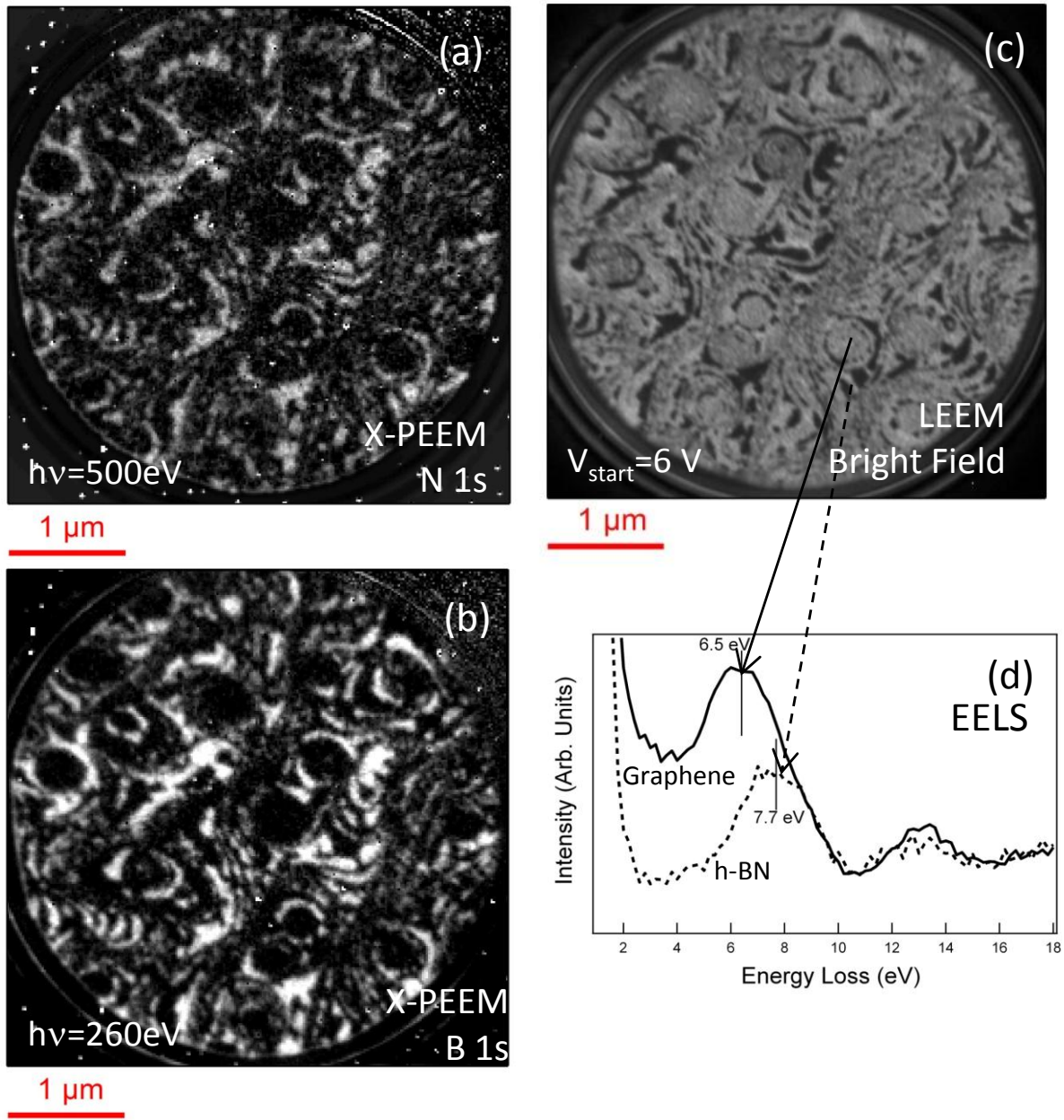

**Fig. S2.** (a) Nitrogen and (b) boron specific XPEEM images and (c) bright field (BF) LEEM image after vacuum vapour deposition of 1 ML Fe on top of h-BNG/Pt(111) and subsequent annealing at 800 K for 5 minutes. (e) μ-EELS spectra obtained integrating the signal on a bright and dark region of the BF LEEM image reveal the collective excitations of the electrons for graphene (6.5 eV) and hexagonal boron nitride (7.7 eV).

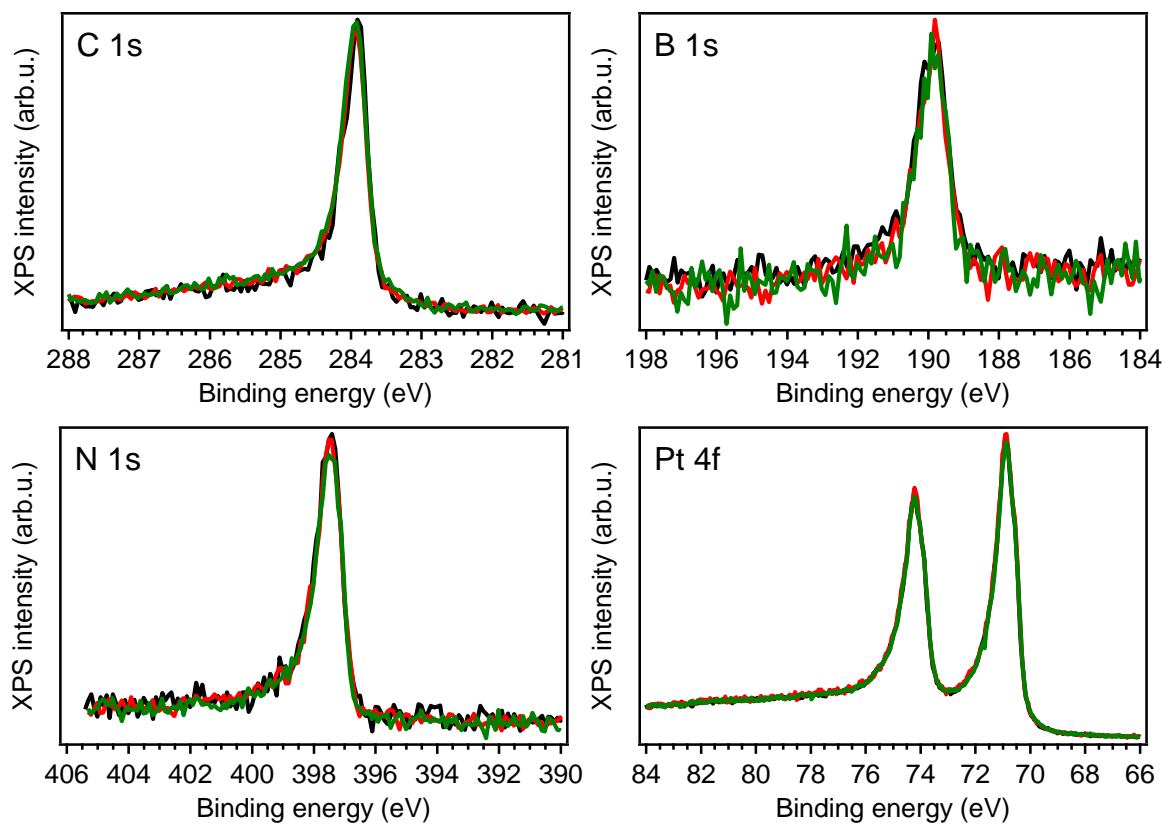

**Fig. S3.** Photoemission core levels of *h*-BNG layer and Pt(111) substrate after the *h*-BNG growth (black curves); subsequent exposure to 100 L of $O_2$ ($p(O_2) = 6.65\times10^{-7}$ mbar) at room temperature (red curves) and at 573 K (green curves). No signs of oxidation or oxygen intercalation are observed under these conditions.

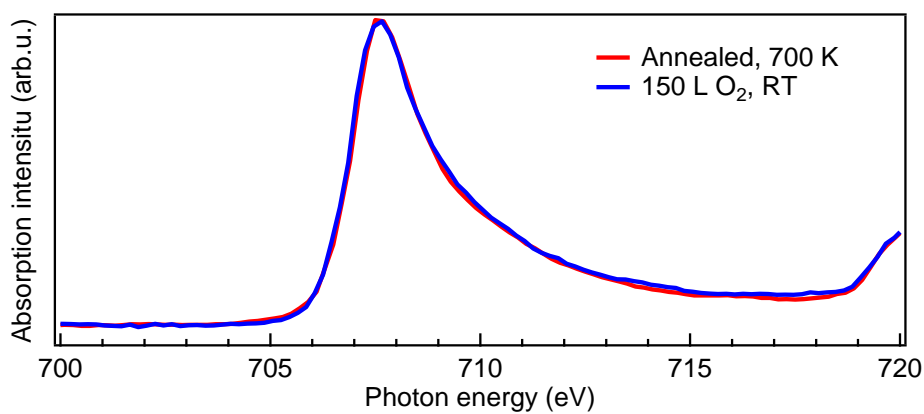

**Fig. S4.** Fe $L_3$-edge TEY XAS of 1 ML Fe deposited on h-BNG/Pt(111) in low pressure $O_2$ atmosphere ($p(O_2) = 1.2\times10^{-8}$ mbar) followed by annealing to 700 K (red line). The sample was then exposed to 150 L $O_2$ ($p(O_2) = 6.65\times10^{-7}$ mbar) at room temperature (blue line).